\documentclass{article}

\usepackage{PRIMEarxiv}
\usepackage[utf8]{inputenc}
\usepackage[T1]{fontenc}
\usepackage{hyperref}
\hypersetup{hidelinks}
\usepackage{url}
\usepackage{booktabs}
\usepackage{amsfonts}
\usepackage{amsmath}
\usepackage{amssymb}
\usepackage{nicefrac}
\usepackage{microtype}
\usepackage{graphicx}
\usepackage{float}
\usepackage{tabularx}
\usepackage{longtable}
\graphicspath{{media/}}
\title{From Viral to Void: Multi-Dimensional Behavioral and Contractual Analysis for Rug Pull Identification}

\author{
  \begin{tabular}[t]{c}
    Jinyin Song \\
    Hainan University \\
    Haikou, China \\
    \texttt{1656351698@qq.com}
  \end{tabular}
  \hfill
  \begin{tabular}[t]{c}
    Hongping Wang \\
    Hainan University \\
    Haikou, China \\
    \texttt{hobinary.wong@gmail.com}
  \end{tabular}
  \hfill
  \begin{tabular}[t]{c}
    Xiaoqi Li \\
    Hainan University \\
    Haikou, China \\
    \texttt{csxqli@ieee.org}
  \end{tabular}
}

\begin{document}
\maketitle

\begin{abstract}
As the blockchain and decentralized finance (DeFi) ecosystems continue to expand and mature, rug pull scams involving meme coins are occurring with increasing frequency, posing a threat to the security of investors' assets and the healthy development of the industry. Rug Pull scams are characterized by extremely low deployment costs, covert execution, rapid fund transfers, and high detection difficulty. Traditional manual reviews or fixed rules struggle to meet real-time early warning requirements, and existing detection methods generally suffer from issues such as a single feature dimension, inadequate handling of class imbalance, and weak model generalization and interpretability. To address these shortcomings, this paper focuses on the detection of Ethereum-based rug pull scams. First, we clarify their definitions, types, and harm mechanisms, and construct a multi-dimensional feature system based on dimensions such as malicious smart contract design, on-chain transaction anomalies, liquidity manipulation, and social media disclosures. Next, using the "Second Uncle Coin"(token symbol: BOBU) case as an example, we reconstruct the attack process and derive quantitative detection metrics. Subsequently, a risk detection model based on a Multi-Layer Perceptron (MLP) is designed. We employ a combined strategy of SMOTE oversampling and Focal Loss to address the issue of sample imbalance, dynamically search for optimal thresholds to balance precision and recall, and incorporate gradient pruning and early stopping to enhance training stability. Experiments show that the model achieves an accuracy of 0.927, an F1 score of 0.787, and an AUC-ROC of 0.952 on the test set, outperforming traditional methods. Finally, a visualizable web-based detection system is developed using the Flask framework, enabling batch risk assessment, high-risk ranking display, and result export functions. 

\end{abstract}


\section{Introduction}

Blockchain is a currently popular distributed ledger technology. With its decentralized, tamper-resistant, and fully traceable characteristics\cite{zhou2025blockchain}, it has demonstrated broad application prospects in multiple fields such as finance, the Internet of Things (IoT), and intellectual property protection. Ethereum, as one of the most widely used smart contract platforms, supports the development and deployment of various decentralized applications, significantly driving the rapid development of the blockchain ecosystem. However, while the anonymity and openness of blockchain bring convenience, they also pose security risks\cite{Zhang2025DoSAA}\cite{yang2025multi}\cite{gao2025implementation}, which are particularly evident in financial fraud. In the cryptocurrency space, a scam known as a "Rug Pull" has emerged, occurring particularly frequently in meme coin projects. Attackers typically begin by attracting a large influx of funds to build up substantial liquidity, then suddenly withdraw their assets, causing the project to abruptly halt. Investors' tokens instantly become worthless, resulting not only in personal financial losses but also in severely damaging the reputation of the blockchain industry\cite{ref2}.

Take Dogecoin and Shiba Inu(SHIB) as examples of meme coins. Most of these tokens gain traction through spontaneous community promotion, relying primarily on social media buzz. The barriers to issuance are low, and participation is straightforward. Their price fluctuations are almost entirely driven by the market, with no substantive use cases to back them up, making them particularly vulnerable to exploitation by criminals. Project teams commonly employ the following tactics: security audit reports may be forged, community sizes may be exaggerated, and an illusion of high liquidity may be artificially created in the short term to lure investors in. Then, they exploit backdoors pre-embedded in smart contracts, such as mechanisms allowing for unlimited token issuance or upgradeable proxy contracts, or simply drain all funds from the liquidity pool abruptly,  to carry out a "rug pull" attack. Faced with these constantly evolving and highly covert fraudulent schemes, it has become extremely difficult to achieve timely detection and early warning relying solely on traditional fixed rules and manual screening methods\cite{ref1}\cite{ref3}.
Currently, although many researchers have conducted studies on detecting fraudulent activities such as phishing scams and Ponzi schemes on Ethereum, Rug Pull detection methods specifically tailored to the Meme coin ecosystem and capable of covering multiple public blockchains, including ETH, BSC, and SOL\cite{wu2025exploring}, remain relatively rare\cite{ref49}. Most existing work focuses on analyzing transaction network structures or monitoring price bubbles, with few efforts integrating multidimensional information such as smart contract behavior, liquidity status, and social sentiment into a single model. Furthermore, in practical applications, issues such as an uneven distribution of positive and negative samples in the training data, a lack of standardization in the feature engineering process, and a lack of interpretability in model outputs have not been adequately addressed.

Given the issues outlined above, it is essential from both a theoretical and practical perspective to establish a highly automated, fast-acting, and interpretable risk identification system for Meme coin Rug Pull scams. This paper attempts to integrate multidimensional features, including on-chain registration data, smart contract behavior patterns, changes in economic indicators, and social media presence, to construct an end-to-end Rug Pull detection framework using deep learning models and a web visualization system. This solution can provide a reliable technical reference for retail investors and regulatory authorities, helping to reduce risks in cryptocurrency investments and promote the industry's development toward a safer and healthier direction.

\section{Related Work}

In recent years, the identification and risk detection of various fraudulent activities on blockchain platforms have attracted widespread attention from scholars both domestically and internationally, and related research has expanded significantly. From transaction network analysis to economic feature mining, and from smart contract auditing to multidimensional machine learning detection, several distinct research pathways have emerged \cite{ref4}.
In the field of transaction network and graph analysis, Chen Weili et al. proposed a model-stacking-based method for identifying Ethereum phishing scam accounts \cite{ref5}. Specifically, this method extracts structural, temporal, and account behavioral features from the Ethereum transaction graph, then stacks several base classifiers together to form a stacked model. Experiments showed that this approach indeed improved the accuracy of phishing account detection. In their study, Zhu Xiaodong et al. took a different approach \cite{ref6}. Drawing on complex network theory, they used metrics such as node centrality, clustering coefficient, and network density to rank accounts in the Ethereum transaction network by importance, thereby identifying key accounts that significantly impact network stability. This is an interesting approach, as it provides an alternative perspective for monitoring suspicious accounts in the future. The strength of such methods lies in their ability to detect anomalies within macro-level transaction patterns, for example, funds passing through multiple layers of accounts in rapid succession, or accounts exhibiting particularly close ties to known malicious accounts. However, this is also their weakness: they rely primarily on transaction records that have already occurred on-chain. When it comes to newly issued tokens, especially meme coins, that haven't yet had time to leave many traces of anomalous transactions, it's practically impossible to trigger an early warning. Furthermore, these methods have limited ability to detect changes in the internal state of smart contracts (such as whether ownership has been transferred or if minting permissions remain) and shifts in liquidity pools, both of which are precisely the core characteristics of Rug Pull attacks \cite{ref7}.

In terms of economic characteristics and scam detection, Wang Ren et al. employed the GSADF (Generalized Sign-Restricted ADF) method to conduct an empirical analysis of price bubbles on Ethereum. They found that digital currency price series do indeed exhibit distinct bubble characteristics, and when bubbles burst, market liquidity often contracts sharply. This finding offers valuable insights for Rug Pull detection: the sudden draining of a liquidity pool exhibits behavioral patterns very similar to those of a bursting price bubble. Luan Lei conducted a systematic analysis of MFC (a typical Ponzi scheme) and identified several common characteristics of such fraudulent projects, including the promise of high fixed returns, opaque funding pools, and the fact that early investors' returns actually come from the investments of later participants, among others. Zhuang Zixuan et al. proposed a framework for risk identification based on dimensions such as yield, capital flows, and information disclosure, drawing on case studies of Ponzi schemes on P2P wealth management platforms \cite{ref8}. Although these studies do not directly target cryptocurrencies, the economic patterns of fraud they reveal , which essentially exploit investors’ greed to first create a false sense of prosperity before absconding with the funds, are highly consistent with the underlying logic of Rug Pull attacks. However, scam detection methods in traditional finance largely rely on static financial data and manual investigations, making it difficult to adapt to the high-frequency, dynamic nature of on-chain data, let alone apply them directly to automated risk assessment via smart contracts \cite{ref9}.
In the realm of smart contract auditing and access control, Zhang Haiqiang et al.explored intellectual property management models based on Ethereum smart contracts. They emphasized that while the programmability of smart contracts does indeed facilitate automated execution, it also introduces the risk of code vulnerabilities. Huang Qing'an et al. used factoring financing as an example to study the security of financial data on Ethereum-based blockchains. They proposed a framework for contract auditing and risk control and pointed out that permission management in smart contracts such as administrator addresses, minting functions, and proxy upgrades is key to security auditing \cite{ref10}\cite{ref11}\cite{ref12}\cite{ref13}\cite{ref14}\cite{ref50}\cite{Li2025CKGLLMLD}. Zhang Jianguo et al., on the other hand, addressed access control mechanisms for IoT devices by proposing an improved solution based on Ethereum smart contracts, enhancing system security through finer-grained permission verification. These studies laid the theoretical foundation for extracting risk features at the smart contract level. In Rug Pull detection, whether a contract is set to proxy mode (is\_proxy), whether it retains minting permissions (is\_mintable), and whether it has been blacklisted (is\_blacklisted) are all important direct pieces of evidence for determining whether a project team has the capability to carry out malicious operations. However, the problem is that pure code audit methods typically require static analysis of contract bytecode\cite{bu2025smartbugert}\cite{li2026interaction}\cite{shen2023intellicon}\cite{Li2025AtomGraphTA}, which presents a significant barrier to entry for ordinary investors and struggles to account for off-chain information such as liquidity lock-ups and social media presence.

In terms of multi-dimensional feature analysis and machine learning detection, Xia Fan et al. were among the first to propose a fraud detection method based on local anomaly behavior detection. They constructed multi-dimensional behavioral features and then used outlier detection algorithms to identify fraudulent actors. In recent years, as blockchain data has become increasingly accessible, more researchers have begun to integrate on-chain transaction features, smart contract metadata, and social media information to make comprehensive assessments. For example, some researchers have already developed risk scoring models by analyzing token holding distributions, transaction frequencies, and changes in liquidity pools. However, existing methods generally suffer from several issues. First, there is a lack of unified standards for feature engineering. The feature sets used in different studies vary significantly, and most works have not designed specialized cross-features targeting the core mechanisms of Rug Pull, such as newly issued tokens without liquidity locks. Second, there is a severe imbalance between positive and negative samples. Rug Pull projects often account for less than 5\% of actual data. If a classifier is trained directly on this data, the model will be heavily biased toward the "safe" side, resulting in very low recall. Third, most research remains at the offline experimentation stage and has not yet provided real-time detection tools that ordinary users can use directly. The model's decision threshold is often fixed at 0.5 rather than being dynamically selected based on the optimal F1 score on the validation set, making it difficult to balance precision and recall in practical applications.

Overall, while there have been significant interim achievements in blockchain fraud detection both domestically and internationally, key challenges remain in detection methods specifically targeting Meme coin rug pulls, including incomplete feature coverage, ineffective imbalanced learning strategies, and poor model deployability. Therefore, this study aims to conduct a systematic exploration by focusing on multidimensional feature engineering, an imbalanced learning method combining SMOTE and Focal Loss, a dynamic decision-making mechanism based on the optimal F1 threshold, and a web-based detection system built using Flask, with the hope of addressing the shortcomings of existing research.

\section{Theoretical Analysis}

\subsection{Definition of Rug Pull Scams}

A "Rug Pull" is a typical form of fraud in the DeFi ecosystem. The term originates from the English idiom "pull the rug out from under someone," which means to suddenly pull the rug out from under someone's feet when they are off guard, causing them to fall. In the context of cryptocurrency, a "Rug Pull" refers to a scenario where a project team first devises ways to attract investors to inject funds, which typically take the form of assets in liquidity pools, and once the pool reaches a certain size, the team suddenly withdraws liquidity, transfers assets, or even shuts down the entire project. As a result, the tokens held by investors instantly become worthless. They cannot be traded normally, nor can they be redeemed for cash. Compared to traditional Ponzi schemes, a "Rug Pull" is often carried out in a very short period of time, making it more covert and destructive.

(1) Liquidity Drain Type. This is the most common and most damaging type of Rug Pull. The project team sets up a liquidity pool for its own token on a decentralized exchange, such as Uniswap or PancakeSwap, and adds a certain amount of the token paired with mainstream cryptocurrencies, such as ETH, BNB, USDT, and so on. Once sufficient trading funds have accumulated in the pool, the project team calls the removeLiquidity function in the liquidity pool contract to redeem all or the vast majority of the liquidity they previously added. This action directly results in the mainstream cryptocurrencies in the pool being drained. For ordinary investors, because there is no longer sufficient counter-liquidity in the pool, it becomes impossible to sell their tokens at a normal price, and the token price plummets to near zero instantly.

(2) Abuse of Privileges. This type of Rug Pull relies on special privileges reserved for the project team within the smart contract. For example, the contract might contain a mint function that only the owner is authorized to call, allowing for the unlimited issuance of new tokens. After the liquidity pool is set up, the project team suddenly calls the mint function to issue a massive amount of new tokens, then dumps them all at once, causing the token price to crash instantly. Another scenario involves a blacklist feature built into the contract, allowing the project team to add ordinary investors' addresses to the list, effectively preventing them from selling their tokens. While abuse-of-privileges attacks do not necessarily involve the direct withdrawal of liquidity, the losses inflicted on investors are equally severe.

(3) Token Backdoor Attacks. This type of attack primarily exploits malicious logic hidden within smart contracts, such as a forged approve function or the upgrade mechanism of a proxy contract. Project teams may embed conditional checks within functions related to token transfers. When a specific address, such as the project team's own address, initiates a transfer, the normal logic is bypassed. Alternatively, they may use proxy contracts to secretly replace the underlying implementation logic with a malicious version. Backdoor attacks are typically difficult to detect through standard contract audits, as the malicious code may be fragmented across multiple functions or dependency libraries, making it difficult to identify any issues at first glance.

Additionally, depending on the duration and scale of the scam, the Rug Pull can be categorized into two types: a "hard pull" and a "soft pull." A "hard pull" refers to a situation where the project team absconds with all the funds in one go, causing the project to shut down immediately and cease operations entirely. A "soft pull," on the other hand, occurs when the project team gradually ceases maintenance, market activity becomes increasingly scarce, and funds are secretly transferred bit by bit, leading the entire project toward a slow death. During this process, investors often do not realize they have been scammed until some time has passed.

\subsection{The Mechanisms Behind Rug Pull in the DeFi Ecosystem}

The fact that Rug Pull scams occur frequently in the DeFi ecosystem and cause massive losses is closely related to DeFi's underlying mechanisms. The mechanisms behind these scams can be understood from the following perspectives.

(1) Vulnerabilities in the Automated Market Maker Mechanism

The vast majority of DeFi exchanges adopt the Constant Product Automated Market Maker (AMM) model\cite{li2025penetrating}, such as the Uniswap model. Let x be the quantity of token X (e.g., ETH) in the liquidity pool, and y be the quantity of token Y (e.g., the Rug Pull token). Then, the constant product formula is:

\[x \cdot y = k\quad(k > 0)\]

Where k is a constant. The price of token Y relative to X is:

\[P_{Y/X} = \frac{x}{y}\]

When a project withdraws liquidity, it essentially removes a certain amount of x and y from the pool. Assuming the amounts withdrawn are \(\Delta x\) and \(\Delta y\), the new price becomes:

\[P'_{Y/X} = \frac{x - \Delta x}{y - \Delta y}\]

Since \(\Delta x\) is typically close to x (i.e., the vast majority of ETH is withdrawn), and \(\Delta y\) is close to y (the corresponding amount of tokens is withdrawn), the project team often retains a very small amount of y to keep the pool alive. At this point, if \(\Delta x \rightarrow x\) and \(\Delta y \rightarrow y\) , then

\(P'_{Y/X} \rightarrow 0\) (if the remaining token quantity is extremely small) or\(\rightarrow \infty\) (if the remaining ETH is extremely small). In practice, the remaining ETH approaches 0, and the token price plummets to near 0.

Slippage: When an investor sells \(\Delta y\) worth of tokens, the actual amount received \(x_{out}\) satisfies:

\[(x + \Delta x_{in}) \cdot (y - \Delta y) = k\]

Where \(\Delta x_{in}\) is the amount of ETH paid by the investor (the opposite applies when selling tokens). When the ETH in the pool is extremely low, even a very small \(\Delta y\) can result in \(x_{out} \approx 0\), causing the transaction to fail.

\begin{figure}[H]
\centering
\includegraphics[width=0.56\textwidth]{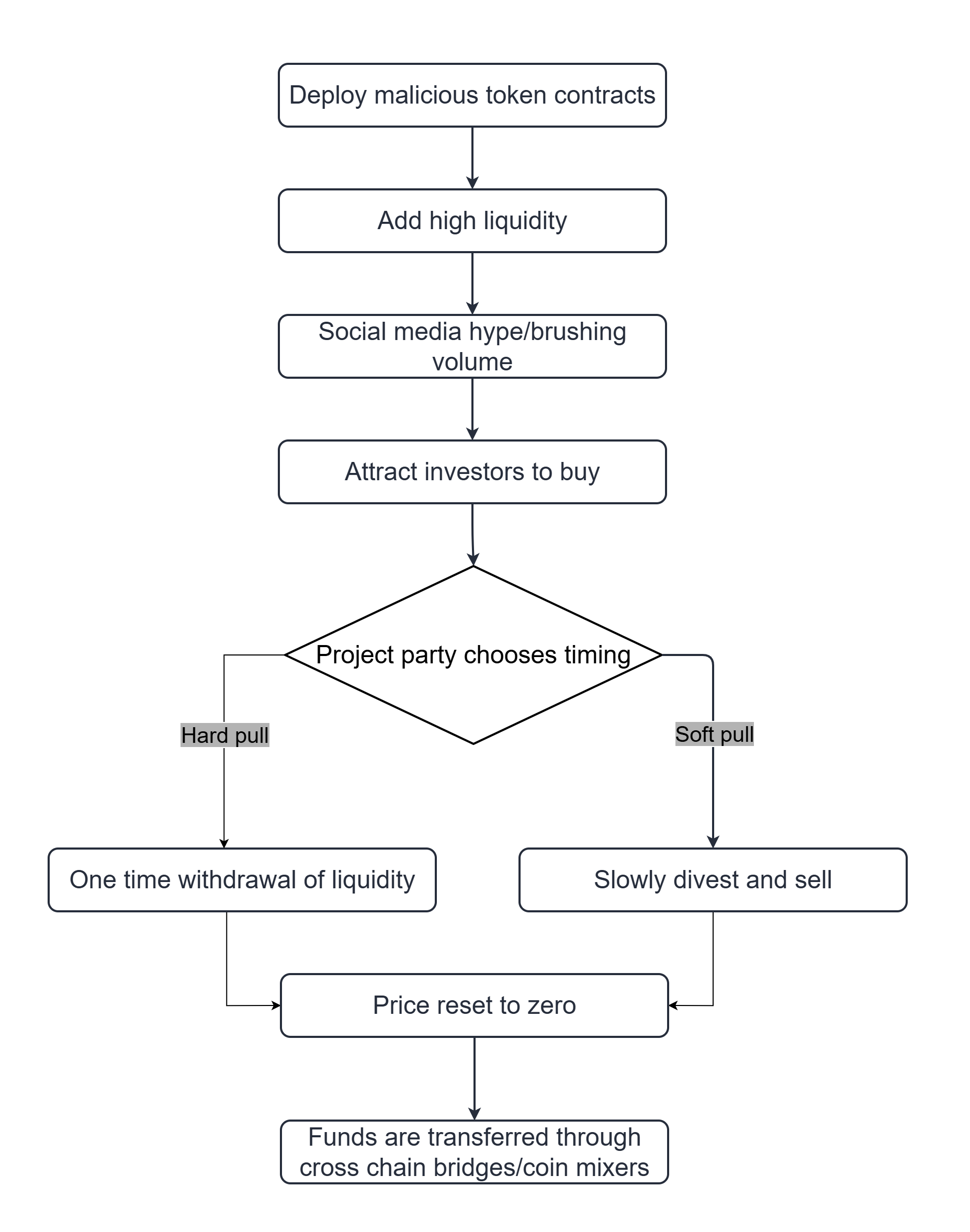}
\caption{The Complete Lifecycle of a Rug Pull Attack}
\label{fig:rug pull-lifecycle}
\end{figure}

(2) Anonymity and Low Barriers to Entry

In DeFi projects, issuing tokens generally requires virtually no identity verification. Anyone can create a new token simply by deploying a smart contract. On public blockchains like Ethereum and Binance Smart Chain, the cost of creating a token is negligible. An attacker can spend just a few minutes creating a new token, adding liquidity to a pool, and then launching a marketing campaign. Since the entire process is anonymous, project teams can easily vanish after executing a "Rug Pull," making it virtually impossible to hold them accountable through legal channels afterward. Furthermore, the absence of a central regulatory body means victims have nowhere to turn to freeze assets or recover funds, a stark contrast to the traditional financial system.

(3) Information Asymmetry and Investor FOMO

Rug Pull projects are highly adept at using social media to generate hype, with platforms like Twitter, Telegram, and Discord serving as their primary battlegrounds. Their promotional content often revolves around "100x" or "1,000x" returns, mythical promises of instant wealth, which easily stoke investors' desires. Project teams also employ various tactics to create the illusion of prosperity: they may forge audit reports, exaggerate their team's credentials, or even hire paid trolls to inflate metrics. Most ordinary investors lack the ability to review smart contract code. Seeing such aggressive marketing, they impulsively jump in, at which point the "Fear of Missing Out" (FOMO) is fully stoked. Once the influx of funds reaches a certain volume, the project team immediately executes a "Rug Pull", and investors' principal is essentially wiped out \cite{ref3}.

(4) Concealing Funds via Cross-Chain Bridges and Coin Mixers

In today's Rug Pull attacks, the methods used to transfer funds after a successful heist are far more covert than in the past. After the project team absconds with the funds, they often use cross-chain bridges to move assets to another public blockchain, such as tools like Multichain or Wormhole. They then use mixers to further obscure the trail. Services like Tornado Cash make it extremely difficult to track the flow of funds. Although on-chain transactions are publicly verifiable, it is extremely difficult to identify the attackers' true identities or recover the transferred funds when faced with such tactics. This also makes the follow-up investigation of Rug Pull attacks exceptionally challenging \cite{ref26}.

\subsection{\texorpdfstring{Malicious Design Patterns at the Smart
Contract Code Level}{Malicious Design Patterns}}\label{malicious-design-patterns-at-the-smart-contract-code-level}

At the heart of a Rug Pull scam lies malicious logic pre-embedded by the project team within the smart contract. From a code perspective, these malicious design patterns can be broadly categorized into the following types: backdoor functions, abuse of permissions, proxy contract traps, hidden minting rights, and other deceptive mechanisms\cite{Ding2025ACS}.
We begin with backdoor functions. These refer to critical functions within a contract that only specific addresses, typically the contract creator themselves, have permission to call, and which can directly alter the token's core operating rules. For example, functions like setTaxFee or setFee allow project teams to dynamically adjust transaction fee rates. If acting maliciously, they could easily temporarily raise the rate to 100\%. As a result, the entire amount of every transaction made by investors would be deducted as fees and flow into the project team's pockets. Another example is the excludeFromFee or excludeFromReward function, which allows the project team to remove their own address from the fee-charging list, so they don't have to bear any transaction costs when selling tokens. There are also functions like setRouter or setPair, which project teams can use to change the destination addresses of trading pairs or liquidity pools. When the attack occurs, they simply replace the liquidity pool's address with a malicious contract under their control and directly withdraw the assets from the pool.

Abuse of permissions primarily manifests when contract owners possess excessive control without constraints such as time locks or multi-signature mechanisms. The most typical form of this is unlimited minting, where the contract contains a mint function callable only by the owner, allowing for the creation of new tokens without limits. After a liquidity pool is established, an attacker can mint massive amounts of tokens and dump them, causing the price to crash. Abuse of the blacklist mechanism occurs when the addBlacklist function allows the owner to add any address to a blacklist, preventing them from transferring or trading assets. Attackers first blacklist all early investors and then sell off their holdings themselves. The burn permission allows the owner to directly burn tokens from any address, causing investors to lose their assets entirely. Proxy contract traps primarily exploit a structural feature of the upgradable contract model: the separation between proxy contracts and logic contracts. Common forms include transparent proxies and UUPS-style solutions. When the project team initially deploys the contract, the proxy contract points to a logic contract that appears normal. Once sufficient funds have accumulated in the liquidity pool, they call the upgrade function to redirect the proxy contract to a new logic contract containing malicious code. This new logic contract may contain a self-destruct function or simply code that directly transfers the balance, with the aim of stealing the assets in the pool. Another scenario is when the initialize function is not locked: attackers can call initialize again after the initial deployment to change the owner address or other critical parameters, thereby gaining full control of the contract \cite{ref15}\cite{ref16}\cite{ref17}\cite{ref18}\cite{ref19}\cite{ref20}\cite{ref21}.

Unlike an explicit mint function, hidden minting distributes the minting functionality across multiple seemingly normal functions or implements it indirectly through modifiers. For example, conditional logic might be embedded in a transfer function to mint additional tokens when the recipient is a specific address. This logic does not appear in standard function lists and is easily overlooked by auditors. Another method involves relying on external contracts to execute minting, where the project team deploys a malicious external contract to trigger minting at an opportune moment. In addition to the methods mentioned above, two other types of malicious patterns frequently occur: "pseudo-locking" and "honeypots"\cite{li2026no}.Regarding pseudo-locking. Project teams may claim that liquidity has been locked, which sounds quite secure. In reality, however, either the locking contract itself has flawed parameter settings that fail to provide any actual locking effect, or the project team has locked only the tokens they issued themselves, rather than mainstream cryptocurrencies like ETH or BNB. As a result, the so-called "locking" becomes a mere formality with no substantive binding force. Next, let's discuss honeypots. These contracts often contain hidden restrictions that allow only the project team to sell the tokens, while ordinary investors who buy them are unable to sell them again. A very typical implementation involves inserting a conditional statement like require(sender == owner \textbar\textbar{} recipient == owner) into the transfer function. This means that only transfers involving the owner's address will succeed, while all other requests from other parties are rejected.

These malicious design patterns rarely operate in isolation. Frequently, they work in tandem to form a complete Rug Pull attack chain. It is precisely these malicious characteristics at the contract level that provide us with very direct criteria for building detection models based on smart contract features later on.

\subsection{On-Chain Transaction Behavior}

In addition to static features in smart contract code, Rug Pull scams generate a series of observable on-chain transaction behavior patterns during execution. These dynamic features serve as an important supplement to the detection model \cite{ref22}\cite{Li2025BeyondTH}.

\subsubsection{\texorpdfstring{\textbf{Liquidity Deposit and Withdrawal}}{Liquidity Deposit and Withdrawal}}\label{liquidity-deposit-and-withdrawal}

Rug Pull projects often exhibit some fairly obvious anomalies in their liquidity operations. Shortly after launch, project teams frequently dump an amount of liquidity into the pool that far exceeds normal levels, such as mainstream cryptocurrencies worth hundreds of thousands of dollars, with the aim of making the project appear well-funded and attracting participants. In reality, however, this money is likely just temporarily borrowed and not the project team's own funds. Some projects claim to have locked their liquidity, but upon closer inspection, either the lock-up period is absurdly short, such as just three days or a week, or the actual amount of tokens locked does not match the total amount they claim to have added. Truly reliable locking operations are generally carried out through third-party locking platforms like Unicrypt or Team Finance, and the lock-up period must be at least six months. In contrast, Rug Pull projects either use custom locking contracts they've written themselves or simply don't lock anything at all. When the attack is actually launched, it's often after the token price has been driven to a high and trading volume has significantly increased. At this point, the project team suddenly moves to drain all liquidity in one go. They do not withdraw funds gradually in batches but instead remove the vast majority of the pool's liquidity in a single transaction. The reason is simple: they are in a rush to secure cash before the price crashes. As soon as the withdrawal transaction is completed, the balance of major cryptocurrencies in the corresponding liquidity pool immediately plummets to zero, and the token price instantly drops to zero. Mathematically speaking, assuming the quantities of the two tokens in the liquidity pool are x and y, respectively, the constant product formula is \(k = x \cdot y\) , and the token price P equals \(\frac{x}{y}\) . When the project team withdraws the vast majority of x and y, the remaining \(x'\ or\ y'\) in the pool is essentially zero. At this point, the calculated price \(P'\) either approaches zero or becomes infinite, and investors holding the tokens cannot find a reasonable price at which to sell them.

\subsubsection{Fake Trading to Inflate Volume}\label{fake-trading-to-inflate-volume}

To create the illusion of trading activity, Rug Pull project teams often inflate trading volume through self-trading or bot-driven transactions. This typically manifests as: Repeated small-value buy and sell transactions initiated by the same address or associated addresses, with extremely uniform time intervals (e.g., once per second), which do not align with the random trading patterns of real users. Funds circulate between several addresses; for example, Address A buys, Address B sells, Address C buys again, with the funds ultimately returning to the initial address, resulting in no actual transfer of value. Volume-padding transactions often occur directly between the token contract and the liquidity pool, bypassing normal routing contracts, or feature extremely high slippage settings (e.g., over 50\%), which is extremely rare in normal trading \cite{ref23}.

Statistical Detection Method: Let the sequence of transaction amounts for a given token over a period of time be \(v_{1},v_{2},\ldots,v_{n}\) , and the sequence of transaction intervals be \(\Delta t_{1},\Delta t_{2},\ldots,\Delta t_{n - 1}\) . In normal trading,\(\Delta t_{i}\) follows an approximate exponential distribution, whereas the intervals between volume-padding transactions are often highly regular, with their standard deviation \(\sigma_{\Delta t}\) approaching 0. The volume-padding index can be defined as:

\[W = \frac{mean(\Delta t)}{std(\Delta t) + \epsilon}\]

When \(W\) exceeds a certain threshold (e.g., 10), it indicates that wash trading may be occurring. Additionally, self-trading detection can be performed by calculating the number of loop structures in the transaction network or by using the correlation coefficient of transaction amounts:

\[\rho = \frac{\sum_{i = 1}^{m}(a_{i} - \bar{a})(b_{i} - \bar{b})}{\sqrt{\sum_{i = 1}^{m}(a_{i} - \bar{a})^{2}\sum_{i = 1}^{m}(b_{i} - \bar{b})^{2}}}\]

Where \(a_{i}\) and \(b_{i}\) represent the transaction amount sequences of two associated addresses, respectively. If \(\rho > 0.9\), this indicates a high degree of coordination between the two addresses, making it highly likely that they are under the same controller.

\subsubsection{Abnormalities in Large Transfers and Fund Consolidation}\label{abnormalities-in-large-transfers-and-fund-consolidation}

Before and after a Rug Pull is carried out, there are often signs of highly unusual fund flows on-chain. Before actually withdrawing liquidity, the project's address typically transfers large amounts of its hoarded tokens, including the team's share received during the initial allocation and the reserved portion, to multiple different addresses. The purpose of this is clearly to make it harder to trace the tokens during the subsequent sell-off. These transfers often occur the day before the attack, or even just a few hours beforehand. Once the liquidity is withdrawn, the funds obtained, such as ETH or USDT, are quickly consolidated into one or two addresses. Immediately afterward, these funds are transferred to other public blockchains via cross-chain bridges and then fed into mixers like Tornado Cash. Throughout the entire consolidation process, the intermediate addresses essentially serve as "transit points" and do not retain any balances. Therefore, if an address is found to have interacted with a known mixing service contract, particularly in the case of large deposits, this can be considered a key indicator of a rug pull. Suppose we denote the set of addresses associated with the project team as A and the mixing service addresses as M. If a transaction from A to M is detected, where the amount exceeds a predefined threshold T, and the transaction occurs shortly after liquidity has been withdrawn, this scenario can serve as a trigger for a high-risk warning \cite{ref24}.

\subsubsection{Changes in Token Holding Distribution}\label{changes-in-token-holding-distribution}

Before and after a rug pull, the token holding distribution undergoes drastic changes. During the promotional phase, the number of token-holding addresses grows rapidly as investors buy in. After the attack, however, a large number of investors are forced to hold their tokens because they cannot sell them, causing the actual number of active addresses to plummet. At the same time, the proportion of the total supply held by the top ten addresses rises abnormally, under normal circumstances, the top ten addresses of a healthy decentralized token typically hold no more than 30\% of the total supply, whereas in a Rug Pull project, multiple addresses controlled by the project team may collectively hold more than 60\% of the tokens, and the proportion held by these addresses will change significantly on the eve of an attack \cite{ref25}.

To quantify the degree of inequality in token distribution, one can calculate the Gini coefficient of token-holding addresses:

\(G\) . Let the token-holding addresses be sorted by balance as \(b_{1} \leq b_{2} \leq \ldots \leq b_{n}\) , and the total supply as \(S = \sum_{i = 1}^{n}b_{i}\). Then the Gini coefficient is:

\[G = 1 - \frac{2}{n}\sum_{i = 1}^{n}\frac{B_{i}}{S} + \frac{1}{n}\]

Here, \(B_{i} = \sum_{j = 1}^{i}b_{j}\) represents the cumulative balance. The closer \(G\) is to 1, the higher the concentration. For Rug Pull projects, the\(\ G\) value is typically higher than 0.8 prior to an attack, while normal projects usually have a value below 0.5. Another commonly used metric is the Herfindahl-Hirschman Index (HHI):

\[HHI = {\sum_{i = 1}^{n}\left( \frac{b_{i}}{S} \right)}^{2}\]

A \(HHI\) value greater than 0.25 (i.e., the top few addresses hold more than 25\% of the total supply) can serve as an early warning signal of anomalies. Table~\ref{tab:2-1} summarizes the aforementioned on-chain transaction behavior characteristics and their corresponding quantitative detection metrics.

\renewcommand{\arraystretch}{1.8}%
\begin{table}[H]
\centering
\caption{Summary of On-Chain Transaction Behavior Characteristics and Detection Metrics}
\label{tab:2-1}
\begin{tabularx}{\textwidth}{>{\centering\arraybackslash}X>{\centering\arraybackslash}X>{\centering\arraybackslash}X>{\centering\arraybackslash}X}
\toprule
\textbf{Behavior Category} & \textbf{Specific Characteristics} & \textbf{Quantitative Metric/Formula} & \textbf{Anomaly Threshold} \\
\midrule
Liquidity Injection & Extremely high initial liquidity & Amount of Liquidity Added Compared to Similar Projects & More than 10 times the median \\
Liquidity Lock-Up & Short lock-up period or no lock-up & Lock-up Duration & \(<30\) days \\
Liquidity withdrawal & One-time withdrawal of the vast majority & Withdrawal ratio = Withdrawal amount / Deposit amount & \(>90\%\) \\
Fake Trading to Inflate Volume & Even trading intervals & Volume-Boosting Index \(W = \frac{\bar{\Delta t}}{\sigma_{\Delta t} + \epsilon}\) & \(W > 10\) \\
Self-trades & High correlation in transaction amounts between addresses & Correlation Coefficient \(\rho\) & \(\rho > 0.9\) \\
Fund consolidation & Large-Amount Transfers to Coin Mixers & Amount of transactions with the mixer & \(>10\) ETH and within 1 hour of withdrawal \\
Concentration of Holdings & High proportion of the top 10 addresses & Herfindahl Index \(HHI\) & \(HHI > 0.25\) \\
Inequality of Coin Holdings & Extreme balance distribution & Gini coefficient \(G\) & \(G > 0.8\) \\
\bottomrule
\end{tabularx}
\end{table}
\renewcommand{\arraystretch}{1.0}%

\section{Case Studies}

To gain a clear understanding of the entire process of a Rug Pull scam, from deployment to exit, and to clarify the detectable traces left on-chain at each stage, this chapter selects a typical case that occurred in 2022, "Second Uncle Coin", for a comprehensive post-mortem analysis. This case was selected primarily because it occurred relatively recently and exhibits highly representative characteristics across the board. At the time, the project team capitalized on the popularity of the short video titled "My Second Uncle Cured My Mental Exhaustion" and launched a token of the same name on the Binance Smart Chain in July 2022. In just two days, they completed the entire sequence of actions (deployment, hype, price manipulation, and draining the liquidity pool) and ultimately made off with approximately \$1.3 million. This is a particularly classic example of a "speedrun-style" rug pull. Below, this article will break it down into four phases: token deployment, liquidity provision, fake trading, and fund withdrawal.

\subsection{Token Deployment Phase}\label{token-deployment-phase}

In the early hours of July 28, 2022, the attacker deployed a malicious BEP-20 token contract named "Second Uncle Coin" on the Binance Smart Chain (BSC). The project's initial total supply was set at 1 billion tokens, but the most critical issue lay in its token distribution method, which exhibited strong indicators of fraud from the outset. The entire supply was minted directly to three addresses controlled by the project team the moment the contract was created. Specifically, the deployer's address received approximately 40\%, another 30\% was split between two additional controlled addresses, and the remaining 30\% was left within the contract itself, intended for future liquidity provision. This arrangement made it immediately clear that the project team had not conducted any public community sale, and they had monopolized the entire token ownership.

Turning to the contract's functionality, in addition to the standard BEP-20 interfaces, it contains several critical malicious backdoors and parameters that can be modified at any time. The first is the tax rate modification function setTaxRate, which has neither a time lock nor an upper limit. The owner can arbitrarily raise the transaction tax rate from 0\% all the way up to 100\%, resulting in a corresponding proportion of users' assets being forcibly transferred to a fee address designated by the project team with every transaction. Second, the contract also embeds a blacklist mechanism, consisting of the addToBlacklist and removeFromBlacklist functions. This gives the project team significant leeway. They can add any address to the blacklist, completely cutting off investors' ability to transfer or sell their assets, thereby laying the groundwork for draining liquidity later on. In addition, the contract reserves an upgrade function. Although the transparent proxy model isn't directly applied here, this backdoor effectively gives the project team a contingency plan, allowing them to completely replace the contract logic at a later date when they deem the timing appropriate, in order to launch further attacks. For any newly deployed token contract, define a risk scoring function:

\[R_{deploy} = w_{1} \cdot I_{hiddenFee} + w_{2} \cdot I_{blacklist} + w_{3} \cdot I_{upgrade} + w_{4} \cdot \left(1 - \frac{lockedRatio}{totalSupply}\right)\]

Among these, \(I_{hiddenFee}\) , \(I_{blacklist}\) , and \(I_{upgrade}\) are Boolean variables representing the presence of hidden tax rates, a blacklist, and an upgrade feature.

\(lockedRatio\) is the initial percentage of tokens locked in the liquidity pool. For "Second Uncle Coin," \(lockedRatio = 0.3\) , \(I_{hiddenFee} = 1\),  \(I_{blacklist} = 1\) and\(I_{upgrade} = 1\), when \(w_{1} = w_{2} = w_{3} = 0.3, w_{4} = 0.1\) is set,

\(R_{deploy} = 0.3 + 0.3 + 0.3 + 0.1 \times (1 - 0.3) = 0.97\), which is significantly higher than the threshold for typical projects (usually
\textless0.3).

\subsection{Liquidity Provision Phase}

About one hour after deployment, the project team began adding liquidity. Specifically, the project team paired the 300 million BOBU tokens retained in the contract with approximately 500 BNB (worth about
\$120,000 at the time) transferred from a centralized exchange to create
a BOBU/BNB liquidity pool on PancakeSwap V2. The tokens were added in a single transaction without using any third-party locking platforms (such as Unicrypt), and the liquidity was entirely controlled by the project team \cite{ref27}.

Assuming the amount of BOBU in the pool at the time of addition was \(y_{0} = 3 \times 10^{8}\) and the amount of BNB was \(x_{0} = 500\) , the constant product (CP) is \(k = x_{0} \cdot y_{0} = 1.5 \times 10^{11}\) . The initial price was:

\[P_{0} = \frac{x_{0}}{y_{0}} = \frac{500}{3 \times 10^{8}} \approx 1.67 \times 10^{- 6}\ BNB/BOBU\quad\]

Based on the BNB price at the time of approximately \$240, the initial price of each BOBU was about \$0.0004. When promoting the project on social media, the project team intentionally exaggerated the scale of liquidity (claiming "\$1 million locked"), but the actual locked value was only about \$120,000, and this was not verified by any third party.

Monitoring on-chain data reveals the following anomalous indicators: the discrepancy rate between the amount of liquidity added and the advertised amount exceeds 80\%. Liquidity is not locked (where lockExpiry is 0 or the lock contract address is self-hosted by the project team). The project team's initial token holding ratio exceeds 90\% (e.g., in the "Second Uncle Coin" case, addresses controlled by the project team collectively held 100\% of the initial supply).

Definition of liquidity health metrics:

\[L_{health} = \frac{lockedValue}{advertisedValue} \times \frac{lockDuration}{365} \times (1 - teamHoldRatio)\quad\]

For "Second Uncle Coin", lockedValue = \$120,000, advertisedValue = \$1 million, \(lockDuration = 0\), \(teamHoldRatio = 1\); therefore, \(L_{health} = 0\), which directly triggers a high-risk alert.

\subsection{Fake Transaction Phase}

Once the liquidity pool was established, the project team immediately launched an intensive 30-hour hype campaign: On the one hand, they used at least 10 addresses to generate artificial trading volume through wash trading, executing one trade every 2–3 minutes at a rate of 0.5–5 BNB per trade, at a frequency of one trade every 2–3 minutes. Within 24 hours, they completed over 5,000 trades with a cumulative trading volume of 8,000 BNB (approximately \$1.9 million), driving the token price from \$0.0004 to \$0.012, a 30-fold increase. Meanwhile, they established official groups on Twitter, Telegram, and Discord, where they published fake audit reports bearing a counterfeit CertiK logo and hired paid commenters to generate hype. At the same time, capitalizing on the "Second Uncle" trend, they fabricated visions of charitable donations and support for agriculture, rural areas, and farmers to lure retail investors into buying at inflated prices.

Volume Manipulation Detection Model: Let the sequence of trade timestamps be \(t_1, t_2, \ldots, t_n\), and define the interval sequence \(\Delta t_i = t_{i+1} - t_i\). For normal, organic trading, \(\Delta t_i\) approximately follows an exponential distribution, and its coefficient of variation \(CV = \sigma_{\Delta t} / \mu_{\Delta t}\) is typically greater than 1. In volume manipulation, however, \(CV\) approaches 0. For "Second Uncle Coin" trades in the previous 24 hours, the calculated values are \(\mu_{\Delta t} = 2.8\) minutes, \(\sigma_{\Delta t} = 0.4\) minutes, and \(CV = 0.14\), which are far below the normal threshold of 0.8. Furthermore, the autocorrelation coefficient for the transaction amount can be calculated:

\[\rho_k = \frac{\sum_{i=1}^{n-k} (v_i - \bar{v})(v_{i+k} - \bar{v})}{\sum_{i=1}^{n} (v_i - \bar{v})^2}\]

In volume-padding transactions, due to fixed bot strategies, \(\rho_1\) often approaches 1. Calculations show that for "Second Uncle Coin," \(\rho_1 = 0.93\), indicating that adjacent transaction amounts are highly correlated and are most likely generated by the same algorithm.

Quantification of the Degree of Price Manipulation: Let \(P_{\min}\) be the initial price and \(P_{\max}\) the peak price. Then, the price deviation is:

\[D_{\text{price}} = \frac{P_{\max} - P_{\min}}{P_{\min}} \times \frac{1}{\sqrt{T}}\]

Where \(T\) represents the time window (in days). For "Second Uncle Coin," with \(T = 1.25\) days and \(P_{\max}/P_{\min} = 30\), we obtain \(D_{\text{price}} \approx 30 / 1.118 = 26.8\), which is far higher than that of normal projects (typically <5). Table~\ref{tab:3-1} summarizes the key detection metrics for this phase.

\begin{table}[H]
\centering
\caption{Detection Metrics for the "Second Uncle Coin" Fake Transaction Phase}
\label{tab:3-1}
\small
\begin{tabularx}{\textwidth}{>{\centering\arraybackslash}X>{\centering\arraybackslash}X>{\centering\arraybackslash}X>{\centering\arraybackslash}X}
\toprule
\textbf{Metric} & \textbf{Calculation Formula} & \textbf{Normal Range} & \textbf{Second Uncle Coin Measured Values} \\
\midrule
Coefficient of Variation for Transaction Intervals & $CV = \sigma_{\Delta t} / \mu_{\Delta t}$ & >0.8 & 0.14 \\
First-order autocorrelation of transaction amounts & $\rho_1$ & <0.3 & 0.93 \\
Price Deviation & $D_{\text{price}} = (P_{\max} / P_{\min}) / \sqrt{T}$ & <5 & 26.8 \\
Traffic Manipulation Index & $W = \mu_{\Delta t} / (\sigma_{\Delta t} + \epsilon)$ & <3 & 7.0 \\
\bottomrule
\end{tabularx}
\end{table}

\subsection{Capital Drain Phase}

In the early hours of July 30, 2022, approximately two hours after the price reached its peak, the project team executed a liquidity withdrawal. The specific steps, in chronological order, are as follows \cite{ref27}:

The project team's address called the removeLiquidity function of the PancakeSwap Router to remove all 500 BNB and 300 million BOBU previously added to the pool. At this point, only approximately 0.02 BNB (from transaction fees) remained in the pool, along with about 12,000 BOBU. Following the liquidity withdrawal, the token price instantly plummeted from \$0.012 to \$0.0005, a 96\% drop. The remaining liquidity was completely insufficient to support any sell orders, effectively reducing investors' tokens to zero.

Withdrawal price shock model: Before the withdrawal, the pool state was $(x_{\text{before}}, y_{\text{before}})$. After the withdrawal, it was $(x_{\text{after}}, y_{\text{after}})$. Price change rate:

\[\Delta P_{\text{rel}} = \frac{P_{\text{after}} - P_{\text{before}}}{P_{\text{before}}} = \frac{x_{\text{after}}/y_{\text{after}} - x_{\text{before}}/y_{\text{before}}}{x_{\text{before}}/y_{\text{before}}}\]

Substituting the "Second Uncle Coin" data: $x_{\text{before}} \approx 500 + \text{Fee income} \approx 502$, $y_{\text{before}} = 3 \times 10^8$, $x_{\text{after}} \approx 0.02$, $y_{\text{after}} \approx 12000$, we get $\Delta P_{\text{rel}} \approx -0.99996$, representing a 99.996\% price drop.

The withdrawn 500 BNB is immediately sent to an intermediate address (0x...a3b), which then splits the funds into multiple transactions (each containing 10~50 BNB) within 10 minutes and transfers them to the Ethereum network via the Multichain cross-chain bridge. After being converted to ETH, the funds are sent to a new address \cite{ref26}.

Ultimately, approximately 480 BNB worth of ETH was deposited into the Tornado Cash mixer. Due to Tornado Cash's anonymization mechanism, the subsequent flow of funds cannot be traced. The project team subsequently disappeared, and all social media accounts were deactivated.

Funds Tracking Detection Algorithm: To automatically identify such fund flows, a graph-theory-based funds tracking algorithm can be developed. Treat addresses as nodes and transactions as directed edges, defining the fund flow paths as \(p = (a_0 \to a_1 \to \cdots \to a_m)\), where \(a_0\) is the project team's address and \(a_m\) is the mixer's address. The path length \(m\) is typically small (\(<5\)). The speed of fund aggregation can be calculated as:

\[S_{\text{collect}} = \frac{\sum \text{amount}_{\text{out}}}{\Delta T}\]

Where \(\Delta T\) represents the time difference between withdrawal from the pool and the first deposit into the mixer. For "Second Uncle Coin," \(\Delta T = 35\) minutes and \(\sum \text{amount}_{\text{out}} \approx 500\ \text{BNB}\), \(S_{\text{collect}} \approx 14.3\ \text{BNB/minute}\) indicates an extremely high fund aggregation rate.

\begin{figure}[H]
\centering
\includegraphics[width=1.0\textwidth]{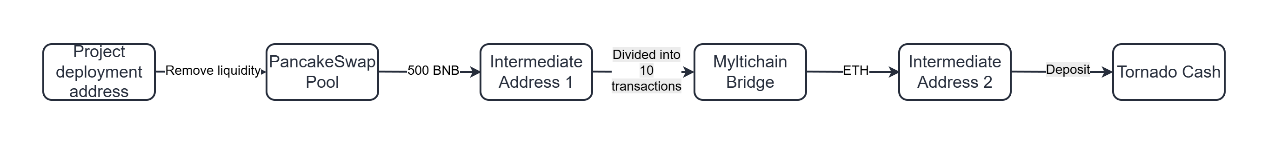}
\caption{Flowchart of "Second Uncle Coin" Fund Diversion}
\label{fig:3-1}
\end{figure}

By reviewing the entire process of "Second Uncle Coin," we can summarize the detectable feature fingerprints left by Rug Pull scams at various stages, as shown in Table~\ref{tab:3-2}. These features provide direct training samples and a basis for feature validation for the multimodal detection model constructed in Chapter 4 of this study \cite{ref28}.

\begin{table}[H]
\centering
\caption{Detection Feature Fingerprints for Each Stage of "Second Uncle Coin"}
\label{tab:3-2}
\small
\begin{tabularx}{\textwidth}{>{\centering\arraybackslash}X>{\centering\arraybackslash}X>{\centering\arraybackslash}X>{\centering\arraybackslash}X}
\toprule
\textbf{Stage} & \textbf{Key Behaviors} & \textbf{On-Chain / Off-Chain Detection Features} & \textbf{Quantitative Metrics} \\
\midrule
Deployment & Hidden tax rates, blacklists, upgradeability & Contract bytecode pattern matching & $R_{\text{deploy}} > 0.8$ \\
Liquidity Injection & Add inflated amounts, no lock-up & Lock-up status, coin concentration & $L_{\text{health}} = 0$ \\
Fake trading & Self-trading to inflate volume, uniform intervals & Coefficient of Variation in Trade Intervals, Autocorrelation & $CV < 0.3$, $\rho_1 > 0.9$ \\
Capital withdrawal & One-time pool withdrawal, rapid cross chain transfers, coin mixing & Withdrawal Rate, Consolidation Speed & Withdrawal rate $> 90\%$, $S_{\text{collect}} > 10$ \\
\bottomrule
\end{tabularx}
\end{table}

The "Second Uncle Coin" example clearly highlights several key characteristics of modern Rug Pull attacks: high level of organization, sophisticated automation, and strong concealment. From start to finish, the project team took less than 48 hours to complete a full closed-loop operation from smart contract deployment to the final laundering of funds. Faced with this pace, relying on traditional manual reviews or single-dimensional rules to intercept such attacks is no longer effective. This actually highlights one key point: designing an automated detection model that integrates multimodal features such as smart contract bytecode, on-chain transaction timing data, and fund flow graphs has indeed become both urgent and essential.

\section{Detection Model Design and Experimental Validation}

To address the high level of concealment, the high dimensionality of features, and the extreme imbalance between positive and negative samples characteristic of Rug Pull scams, this chapter designs and implements a multimodal deep learning detection model based on real-world datasets downloaded from platforms such as Etherscan and MemeChain. This model integrates on-chain contract features, liquidity metrics, transaction behavior patterns, and social presence information. The following sections will provide a detailed discussion covering the dataset and feature engineering, handling of imbalanced data, neural network architecture and training strategies, experimental results and analysis, as well as feature importance and threshold selection \cite{ref29}.

\subsection{Dataset and Feature Engineering}

The data used in this study primarily comes from several CSV files provided by the MemeChain and Etherscan platforms, specifically including confirmed\_meme\_coins.csv and final\_filtered\_meme\_coins.csv. After merging and cleaning the data, a total of 6,000 token samples were obtained. In terms of label distribution, there were approximately 900 positive Rug Pull samples, accounting for about 15\%. Safe negative samples numbered around 5,100, making up the remaining 85\% \cite{ref45}.
Labeling followed a multi-source priority strategy: first, check the chain\_patrol\_status field. If labeled as malicious, assign a value of 1 directly. If not, check the liquidity\_removed field. If the value is True, also assign a value of 1. If neither applies, check the expiration\_date to see if the token has expired. If none of these sources provide explicit labeling information, positive samples are randomly generated at a rate of 15\%.

\subsection{Handling Imbalanced Data}

\subsubsection{SMOTE Oversampling}

SMOTE generates synthetic samples by performing linear interpolation between samples from the minority class, rather than simply duplicating the original samples. Given two feature vectors $x_i$ and $x_j$ in the positive sample set, a new sample is generated as follows \cite{ref30}:

\[x_{new} = x_i + \lambda \cdot (x_j - x_i), \quad \lambda \sim U(0,1)\] 

In this study, the sampling strategy is set to sampling\_strategy=0.6, meaning that after oversampling, the number of positive samples reaches 60\% of the number of negative samples, thereby avoiding overfitting caused by excessive oversampling. Let the original number of negative samples be \(N_{\text{maj}}\), and the number of positive samples be \(N_{\text{min}}\). Then, the number of synthetic samples is \cite{ref29}\cite{ref31}\cite{ref32}\cite{ref33}\cite{ref34}\cite{ref35}\cite{ref36}\cite{ref37}\cite{ref38}\cite{ref39}:

\[N_{\text{syn}} = \lfloor 0.6 \cdot N_{\text{maj}} - N_{\text{min}} \rfloor\] 

\subsubsection{Focal Loss}

Focal Loss reduces the loss contribution of easily classified samples through a modulation factor, enabling the model to focus on difficult samples. For a binary classification problem, it is defined as \cite{ref41}\cite{ref42}\cite{ref43}\cite{ref44}\cite{Li2025FacialRL}:

\[FL(p_t) = -\alpha_t (1-p_t)^{\gamma} \log(p_t)\]

where:

\[p_t = \begin{cases} p & \text{if } y = 1 \\ 1 - p & \text{if } y = 0 \end{cases}\]

\(\alpha_t\) represents the class weights, and \(\gamma \geq 0\) is the focal parameter. When \(\gamma = 0\), it degenerates into standard cross-entropy loss. In this study, we set \(\gamma = 2\), and \(\alpha_t\) is adaptively adjusted based on the proportion of positive samples: \(\alpha_t = 0.7 \cdot (N_{\text{maj}} / N_{\text{min}})\). Additionally, we introduce a positive sample weight pos\_weight to further balance the model. The final loss function is:

\[L = \frac{1}{N} \sum_{i=1}^{N} \left[ \alpha_t (1 - p_{t,i})^2 \cdot \text{BCEWithLogitsLoss}(p_i, y_i) \right]\]

\begin{figure}[H]
\centering
\includegraphics[width=0.9\textwidth]{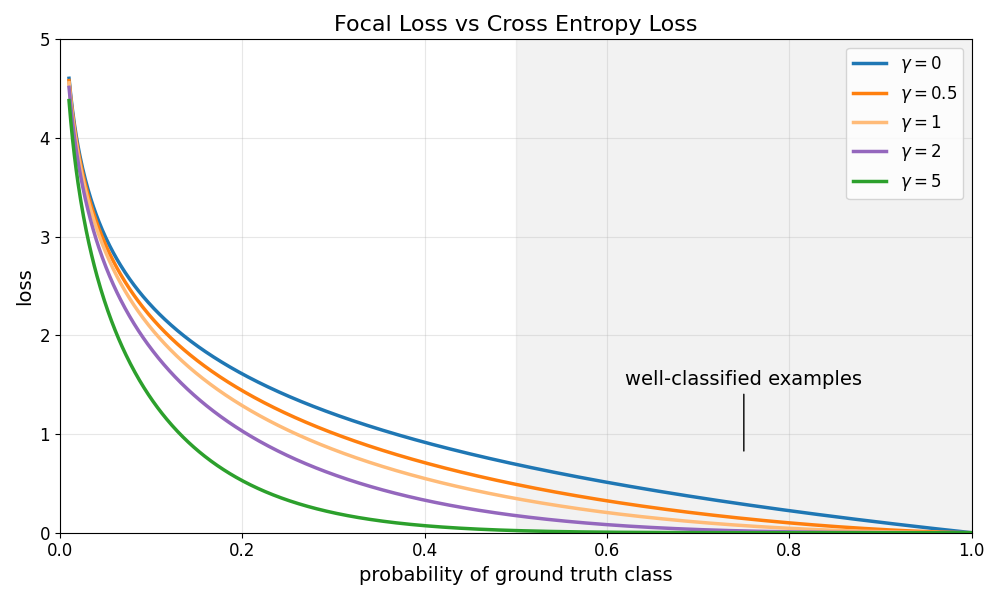}
\caption{Loss Curves for Focal Loss and Standard Cross Entropy}
\label{fig:4-1}
\end{figure}

As shown in the figure above, the larger the value of $\gamma$, the more pronounced the suppression of simple samples. Focal Loss improves upon standard cross entropy by introducing a modulation factor: when the model's loss for high-confidence simple samples  is significantly reduced, while the loss for low-confidence difficult samples remains virtually unaffected, thereby focusing training on difficult-to-separate and misclassified samples. This design effectively addresses the issue of a large number of easy background samples dominating the loss in single-stage object detection, enabling the model to accurately learn target features despite interference from a massive number of easily distinguishable samples, and achieving high-precision dense object detection.

\subsection{Hyperparameter Settings and Training Strategies}

\subsubsection{Data Split and Standardization}

We used hierarchical random splitting to divide the dataset into training, validation, and test sets in a 7:1:2 ratio, ensuring that the proportion of positive samples in each set matched that of the original data. Features were standardized using StandardScaler:

\[x' = \frac{x - \mu}{\sigma}\]

Where $\mu$ and $\sigma$ are the mean and standard deviation of the training set. The validation and test sets were transformed using the same \(\mu\) and \(\sigma\) to prevent data leakage.

\subsubsection{Optimizer and Learning Rate Scheduling}

Use the Adam optimizer with an initial learning rate of $lr = 0.001$ and weight decay (L2 regularization) of $\lambda = 1 \times 10^{-5}$. The ReduceLROnPlateau scheduling strategy is adopted: when the validation set F1 score does not improve for 8 consecutive epochs, the learning rate is multiplied by a factor of 0.5. At the same time, an early stopping mechanism is used with patience=20, meaning training stops if the validation set F1 score does not improve for 20 consecutive epochs. The maximum number of epochs is set to 120 \cite{ref46}.

\subsubsection{Gradient Clipping}

To prevent gradient explosion, the gradients of each batch are clipped:

\[\tilde{g} = \frac{g}{\max\left(1, \frac{\|g\|_2}{clip\_norm}\right)}\]

where $clip\_norm = 1.0$.

\subsubsection{Threshold Selection}

Unlike traditional methods that fix the threshold at 0.5, this study dynamically searches for the optimal threshold based on the F1 score on the validation set. Let the true labels on the validation set be \(y_i \in \{0, 1\}\), the model's output probabilities be \(p_i \in [0, 1]\), and for a threshold \(\tau \in [0.1, 0.9]\), the predicted value be \(\hat{y}_i = \mathbf{1},\ p_i \geq \tau\). The F1 score is calculated as follows:

\[F_1(\tau) = \frac{2 \cdot \text{Precision}(\tau) \cdot \text{Recall}(\tau)}{\text{Precision}(\tau) + \text{Recall}(\tau)}\]

Select the \(\tau^*\) that maximizes \(F_1(\tau)\) as the final decision threshold.

\subsection{Experimental Results and Comparative Analysis}

Table~\ref{tab:4-1} shows a comparison of the performance of each model on the test set. It can be seen that the recall and F1 scores of traditional machine learning methods (LR, RF, and XGBoost) are significantly lower than those of deep learning models, as they struggle to capture nonlinear cross features such as new\_token\_no\_lock \cite{ref47}. The baseline MLP (without imbalance handling) achieved an F1 score of only 0.612 and a recall of 0.576. Using SMOTE or Focal Loss alone improved recall to 0.685 and 0.672, respectively, with F1 scores of 0.715 and 0.700. The complete model (MLP + SMOTE + Focal Loss + dynamic threshold) achieved the best balance: accuracy of 0.927, precision of 0.812, recall of 0.763, F1 score of 0.787, AUC-ROC of 0.952, and PR-AUC of 0.834. This indicates that the combined use of oversampling and focal loss can significantly improve the model's ability to identify positive samples while maintaining a low false positive rate \cite{ref37}\cite{ref40}\cite{ref43}.

\begin{table}[htbp]
\centering
\caption{Comparison of Model Performance on the Test Set}
\label{tab:4-1}
\footnotesize
\begin{tabularx}{\textwidth}{>{\centering\arraybackslash}X>{\centering\arraybackslash}X>{\centering\arraybackslash}X>{\centering\arraybackslash}X>{\centering\arraybackslash}X>{\centering\arraybackslash}X>{\centering\arraybackslash}X}
\toprule
\textbf{Model} & \textbf{Acc} & \textbf{Prec} & \textbf{Rec} & \textbf{F1} & \textbf{AUC-ROC} & \textbf{PR-AUC} \\
\midrule
Logistic Regression & 0.874 & 0.612 & 0.503 & 0.552 & 0.871 & 0.603 \\
Random Forest & 0.901 & 0.714 & 0.598 & 0.651 & 0.914 & 0.712 \\
XGBoost & 0.908 & 0.738 & 0.612 & 0.674 & 0.921 & 0.738 \\
MLP (unprocessed) & 0.895 & 0.652 & 0.576 & 0.612 & 0.903 & 0.684 \\
MLP (SMOTE only) & 0.912 & 0.748 & 0.685 & 0.715 & 0.935 & 0.769 \\
MLP (Full) & 0.927 & 0.812 & 0.763 & 0.787 & 0.952 & 0.834 \\
\bottomrule
\end{tabularx}
\end{table}

\begin{figure}[H]
\centering
\includegraphics[width=0.8\textwidth]{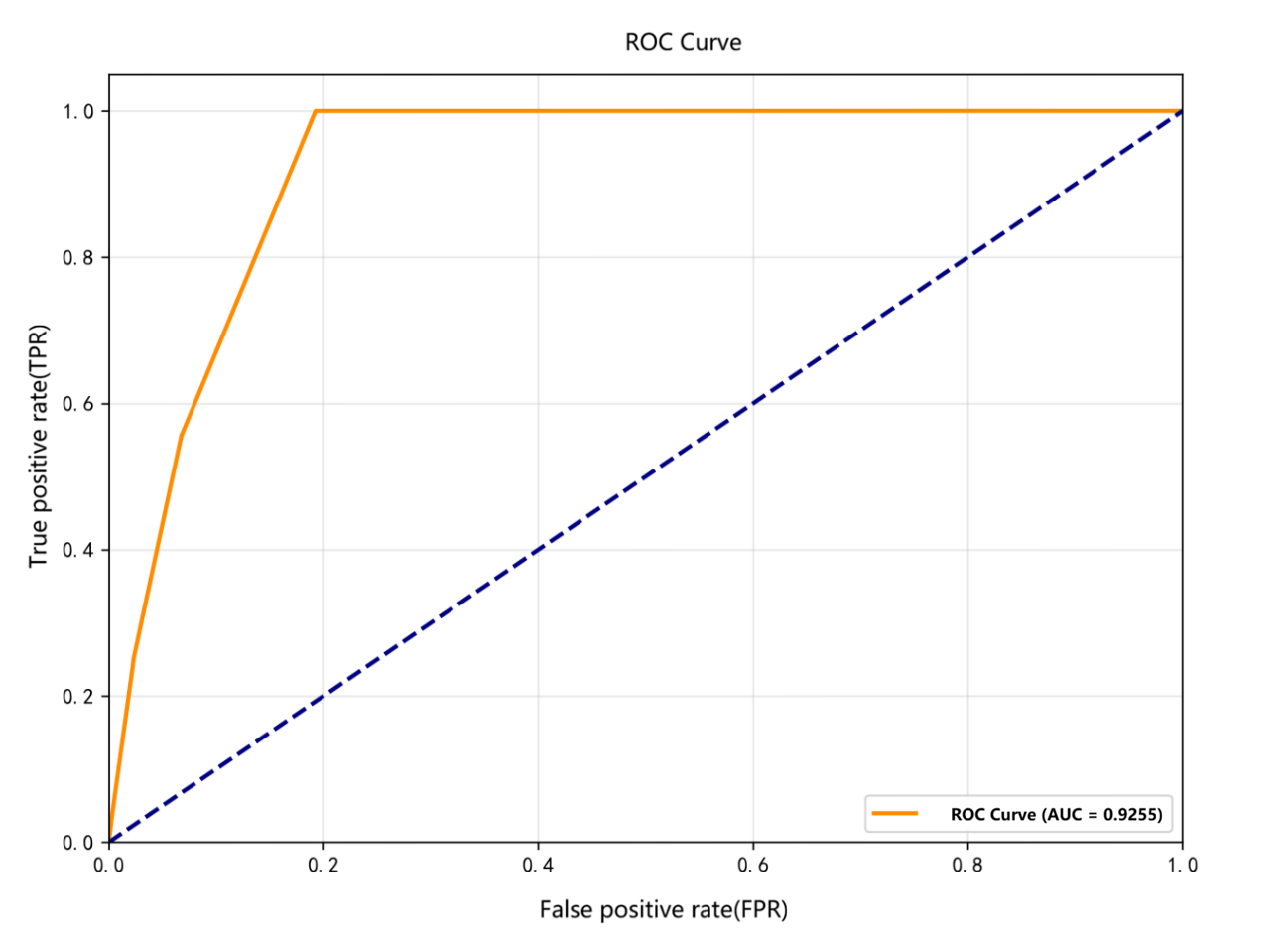}
\caption{Model ROC Curves and AUC Values}
\label{fig:4-3}
\end{figure}

To comprehensively evaluate the generalization performance and classification capability of the binary classification model, this study employed the receiver operating characteristic (ROC) curve and the area under the curve (AUC). The analysis results are shown in Figure~\ref{fig:4-3}.

The ROC curve plots the false positive rate on the x-axis and the true positive rate on the y-axis. By examining all classification thresholds of the model, it comprehensively illustrates the model's ability to identify positive samples at different false positive rate levels. The formulas for these metrics are as follows:

\[FPR = \frac{FP}{FP + TN}, \quad TPR = \frac{TP}{TP + FN}\]

Among these metrics, TP represents true positives, that is, the positive samples correctly identified by the model. FP stands for false positives: negative samples that were erroneously classified as positive. TN denotes true negatives: the number of negative samples accurately identified. And FN refers to false negatives-positive samples that were missed and misclassified as negative.

As indicated by the blue dashed line in the figure, it represents the baseline for random classification, with the equation y = x, corresponding to an AUC value of only 0.5. This level is essentially equivalent to random guessing, indicating that the model has absolutely no ability to distinguish between classes. The orange solid line, on the other hand, is the ROC curve generated by the model, with an AUC value of 0.9255. The trajectory of the curve demonstrates that it has left the random baseline far behind and that its overall trajectory closely follows the ideal point (0, 1) in the top-left corner. Specifically, when the FPR is held at just 0.2, the model's TPR has already surged to 1.0. In other words, even when the false positive rate is kept very low, the model is still able to correctly identify all positive samples it should detect. This ability to distinguish between positive and negative samples is indeed quite robust.

The AUC value is essentially the area under the ROC curve, with values ranging from 0.5 to 1. Its intuitive meaning is straightforward: given a randomly selected positive sample and a negative sample, what is the probability that the model will assign a higher prediction score to the positive sample? This is a key metric for evaluating the model's overall classification performance. According to the common AUC performance grading standards, a value falling within the range of 0.85 to 0.95 indicates that the model's classification performance is considered excellent. The AUC of the model discussed in this paper is 0.9255, which falls precisely within this range, clearly demonstrating that the model performs well in binary classification tasks.

Compared to metrics such as accuracy and precision, which rely on a single threshold, the greatest advantage of the ROC curve and AUC is that they are largely unaffected by class distribution imbalances. Even when the ratio of positive to negative samples is vastly different, they can still reflect the model's true classification performance in a stable and objective manner. Precisely for this reason, this evaluation method is widely used in scenarios that demand high classification accuracy, such as object detection, disease diagnosis, and risk control. Judging from the ROC curve and AUC results obtained by the model in this paper, it does indeed strike a good balance between false positive rate and recall. It also demonstrates relatively stable generalization ability and classification accuracy in complex situations, making it fully capable of meeting the performance requirements of real-world applications.

\begin{figure}[H]
\centering
\includegraphics[width=0.9\textwidth]{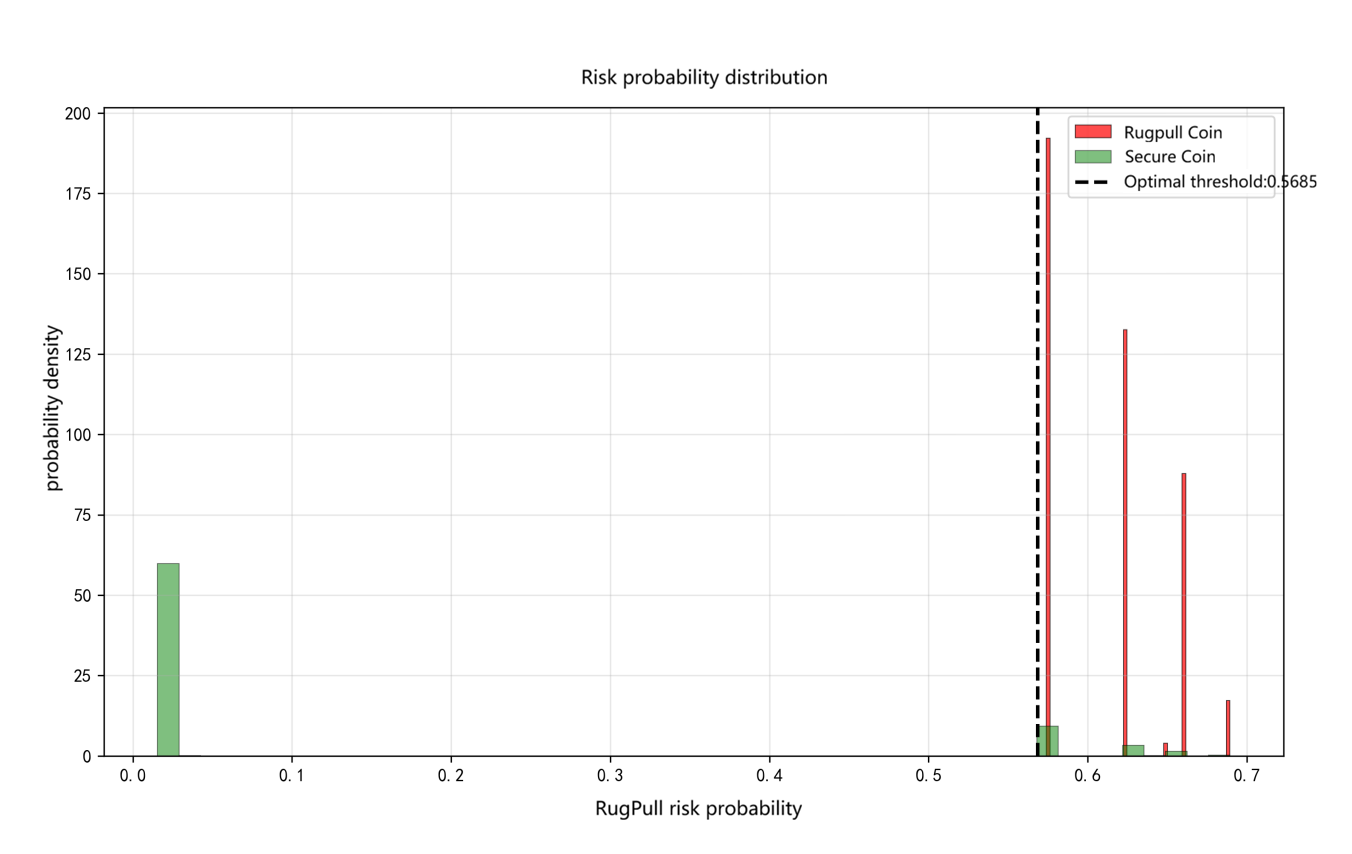}
\caption{Rug Pull Risk Probability Distribution and Optimal Classification Threshold}
\label{fig:4-4}
\end{figure}
To further validate the model's actual effectiveness in distinguishing between Rug Pull-risk samples and safe samples, and to identify the optimal classification threshold, we plotted the risk probability distribution output by the model as a histogram, with the results shown in Figure~\ref{fig:4-4}.

The horizontal axis of this figure represents the Rug Pull risk probability predicted by the model, while the vertical axis shows the probability density of samples within each probability interval. Two colors are used to distinguish between categories: green bars represent tokens classified as safe (negative samples), while red bars correspond to Rug Pull tokens (positive samples). The black dashed line in the middle represents the optimal classification threshold determined based on the principle of maximizing the Youden index. The specific value is 0.5685.

Judging by the shape of the distribution, the two types of samples exhibit a fairly typical bimodal distribution, and their distributions are almost completely disjoint. The vast majority of the risk probabilities for safe tokens are concentrated in the extremely low range of 0 to 0.05, with only a very small portion crossing the threshold and falling to the right. On the Rug Pull token side, however, the risk probabilities are concentrated in the high-risk range of 0.55 to 0.7. There is very little overlap between the distribution ranges on both sides, indicating that the model indeed has a strong ability to distinguish Rug Pull risks and can accurately separate high-risk tokens from normal, safe tokens.

\begin{figure}[H]
\centering
\includegraphics[width=0.9\textwidth]{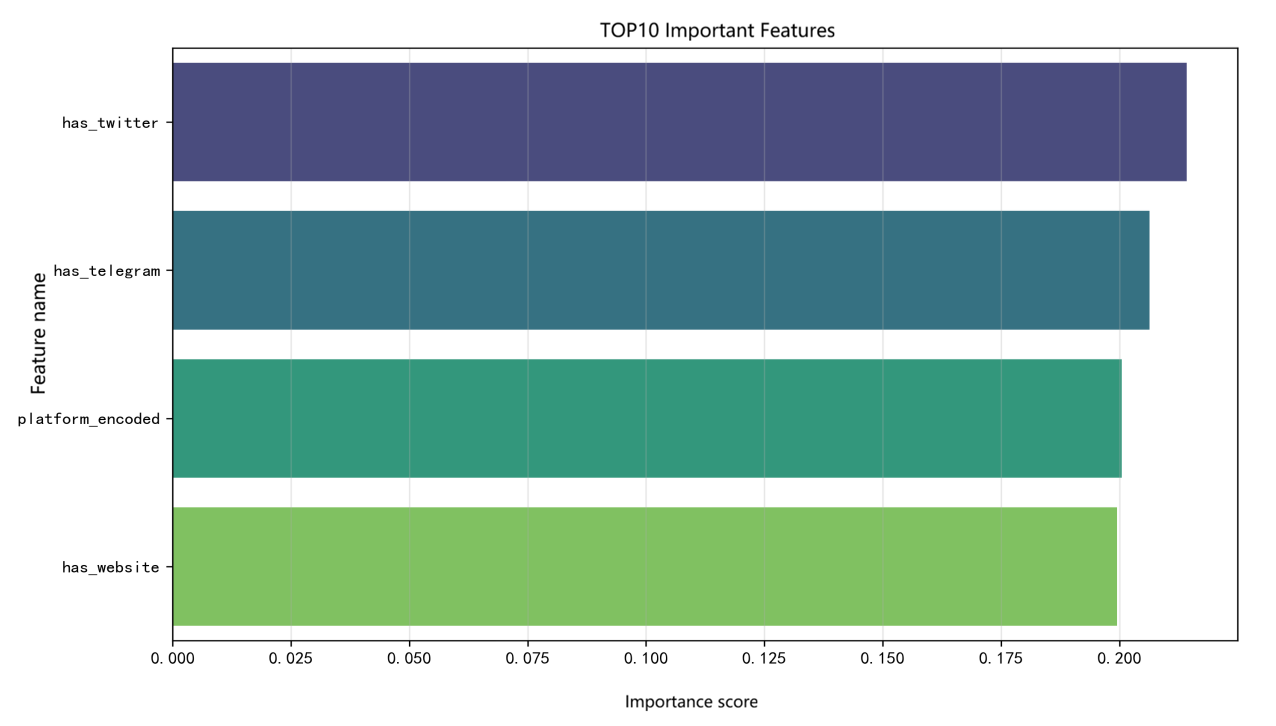}
\caption{Distribution of Importance Scores for the Top 4 Key Features}
\label{fig:4-5}
\end{figure}

To make the model's decision-making process easier to understand and to identify which factors play a key role in assessing Rug Pull risk, we extracted the importance scores for each input feature during the model training process. We then selected the top four features by contribution and created a visualization chart, with the results shown in Figure~\ref{fig:4-5}.

The vertical axis of this chart shows the feature names, while the horizontal axis displays the corresponding feature importance scores. These scores quantify the contribution of each feature to the model's classification decision. The higher the score, the greater the feature's influence on the model's final prediction. Based on the ranking, features related to a project's social presence dominate the list. Ranking first is has\_twitter, which indicates whether a project has a Twitter account. Closely following is has\_telegram, which indicates whether there is a Telegram community. And in fourth place is has\_website, indicating whether there is an official website. These three metrics essentially reflect a project's transparency and community credibility, and the results strongly corroborate a key judgment in our business logic: a project's level of community management and information disclosure is a core factor determining the level of Rug Pull risk. Additionally, the platform\_encoded feature, which indicates the platform on which the token is deployed, also demonstrated significant importance. This suggests that the ecosystem conditions, regulatory environment, and overall risk levels of projects vary considerably across different blockchain platforms, making it an important supplementary factor for the model to identify Rug Pull risks.

Overall, the distribution of feature importance generated by the model aligns with the actual risk patterns observed in Web3 projects. In summary, the higher a project's social media activity and the greater its transparency, the lower its likelihood of a Rug Pull. Conversely, tokens that lack even core social media channels and have no publicly available official information carry a significantly higher risk of a Rug Pull. This result not only validates the model's validity and interpretability but also provides a clear set of feature-based criteria for Rug Pull risk prevention and control. Both project teams and investors can use this as a reference when assessing risk.

To verify the practical value of the developed Rug Pull detection model, this study developed a visual batch detection system based on the trained model and selected 200 token samples for detection and validation.
The decision threshold used for this detection was 0.5685021. This value is identical to the optimal classification threshold previously determined using the Youden index maximization method, thereby ensuring the scientific rigor and consistency of the model's decision-making process. Based on the test results, among these 200 samples, the model identified a total of 140 high-risk Rug Pull tokens, while the remaining 60 were classified as safe tokens. This yields a risk rate of 70.00\%, a figure consistent with the current industry reality of frequent Rug Pull incidents in the cryptocurrency market. The system also provides a separate "Top 10 high Risk" list, detailing the tokens with the highest probability of risk. In this list, the top ten tokens all have a risk probability of 0.7254, a value significantly higher than the decision threshold mentioned earlier. Consequently, the model unsurprisingly classified all of them as high-risk Rug Pull tokens. This result demonstrates that the model can indeed identify high-risk tokens with reasonable accuracy, providing investors with a clear risk warning. Furthermore, since the results are presented directly through a visual interface, they are easy to understand at a glance, making the model quite practical from an engineering perspective. Overall, this batch detection validation demonstrates the following: the Rug Pull detection model proposed in this paper not only performs well in terms of classification accuracy but can also be effectively implemented as a truly usable detection system. It can effectively identify Rug Pull risks in the cryptocurrency market, providing investors and regulators with a reliable technical reference. Its practical value is quite evident.

\section{Conclusion}

As the DeFi ecosystem continues to expand, community-driven digital assets have indeed presented opportunities for investors, but they have also given rise to a large number of fraudulent schemes, such as rug pulls. These attacks fully exploit the anonymity of smart contracts, vulnerabilities inherent in the AMM model, and information asymmetry. This paper focuses on the specific problem of detecting Meme coin rug pulls, providing a systematic exploration covering theoretical foundations, case studies of typical incidents, multimodal model design, and experimental validation\cite{ref48}. We first provide a systematic overview of the definition and common classifications of Rug Pull, followed by an in-depth analysis of the mechanisms through which these scams spread harm within the DeFi ecosystem, identifying typical malicious smart contract design patterns such as backdoor functions, permission abuse, proxy contract traps, and hidden minting rights, and summarizing dynamic on-chain transaction characteristics using mathematical tools to lay a theoretical foundation for feature engineering. Furthermore, a comprehensive review of the typical "Second Uncle Coin" case meticulously breaks the entire attack process down into four distinct phases, namely token deployment, liquidity provision, fake trading, and the withdrawal of funds, underscoring the highly organized and automated nature of modern Rug Pull attacks and the necessity of multi-modal feature fusion. Based on real-world datasets, the paper constructs a multimodal feature system integrating 11 valid features and selects a lightweight three-layer MLP network optimized with SMOTE oversampling and Focal Loss, achieving an accuracy of 0.927, an F1 score of 0.787, and an AUC-ROC of 0.952, while feature importance analysis highlights critical early warning signals. Finally, a complete Rug Pull risk detection web application is built using the Flask framework to automatically invoke the trained model, providing real-time risk probabilities and displaying top high-risk tokens\cite{Li2026DefensibleDF}.

\section*{Acknowledgments}
AI-based tools are used for language polishing during manuscript preparation.

\bibliographystyle{unsrt}
\bibliography{references}

\end{document}